\documentclass[rnote]{aa}  
\usepackage{graphicx}
\usepackage{txfonts}
\newcommand{\pun}[1]{\mbox{\rm\,#1}} 

\newcommand{\draftflag}{false}

\newcommand{\beq}{\begin{equation}}
\newcommand{\eeq}{\end{equation}}

\newcommand{\eref}[1]{\mbox{(\ref{#1})}}

\newcommand{\I}{\ensuremath{I}}
\newcommand{\Irot}{\ensuremath{\tilde{I}}}

\newcommand{\Frot}{\ensuremath{\tilde{F}}}
\newcommand{\vsini}{\ensuremath{V\sin(i)}}
\newcommand{\vvsini}{\ensuremath{V^2\sin^2(i)}}
\newcommand{\vsinimu}{\ensuremath{\tilde{v}}}
\newcommand{\rotint}{\ensuremath{\int^{+\vsinimu}_{-\vsinimu}\!\!d\xi\,}}
\newcommand{\imu}{\ensuremath{m}}
\newcommand{\imupone}{\ensuremath{{m+1}}}
\newcommand{\nmu}{\ensuremath{N_\mu}}
\newcommand{\msum}[1]{\ensuremath{\sum_{#1=1}^{\nmu}}}
\newcommand{\wmu}{\ensuremath{w_\imu}}

\begin{document}
   \title{3D spectral synthesis and rotational line broadening}

\author{ Hans-G\"unter Ludwig\inst{1}}

\institute{GEPI, CIFIST, Observatoire de Paris-Meudon, 5 place Jules Janssen, 92195 Meudon
  Cedex, France\\ \email{Hans.Ludwig@obspm}}

\date{Received ???; accepted ???}

 
  \abstract
  {Spectral synthesis calculations based on three-dimensional stellar atmosphere models are
    limited by the affordable angular resolution of the radiation field. This
    hampers an accurate treatment of rotational line broadening.}
  {We aim to find a treatment of rotational broadening of a spherical star when the
    radiation field is only available at a modest number of limb-angles.}
   {We apply a combination of analytical considerations of the line-broadening
     process and numerical
   tests.}
   {We obtain a method which is closely related to classical flux convolution and
   which performs noticeably better than a previously suggested procedure.
     It can be applied to rigid as well as differential rotation. }
   {}

\maketitle
%

\section{Introduction}

Spectral synthesis calculations based on time-series of three-dimensional (3D)
background structures from hydrodynamical model atmospheres are
computationally demanding and pose limits on the affordable resolution
for representing the angular dependence of the radiation field. In such
calculations, the number of employed azimuthal directions typically ranges
from four to eight, the number of inclined directions representing the
center-to-limb variation being between three and five.  Obviously, the resolution is
not high, and in particular makes an accurate implementation of rotational
line broadening somewhat difficult. To our knowledge, the only published method
handling rotational broadening in the 3D case is  that of Dravins \&\ 
Nordlund (1990). Here, we describe an alternative procedure which provides
higher accuracy at similar computational complexity. It is closely related to
standard flux convolution (e.g., Gray 1992).  Its development was
motivated by the demand for an accurate description of the rotational
broadening of spectral lines in the solar spectrum (Caffau et al. 2007).

\section{The problem and basic assumptions}

For the moment, we neglect differential rotation, and treat the star
  as spherical and rigidly rotating. Local-box hydrodynamical model
atmospheres provide a statistical realization of a small patch of the surface
flow pattern. Formally, we want to obtain an estimate of the expectation value
of a rotationally broadened, disk-integrated line profile. Rotational symmetry
with respect to the stellar disk center implies that there is no azimuthal
dependence of the radiation field. All we need to know for evaluating the
disk-integrated rotationally broadened line profile are temporal and azimuthal
averages of the emergent radiation intensity of the local model as a function
of asterocentric angle $\vartheta$.  Hence, the problem is equivalent to the
rotational line broadening problem in standard, plane-parallel model
atmospheres.

\section{The method of Dravins \&\ Nordlund}

\begin{figure}
\begin{center}
\resizebox{0.75\hsize}{!}{\includegraphics[draft = \draftflag]%
{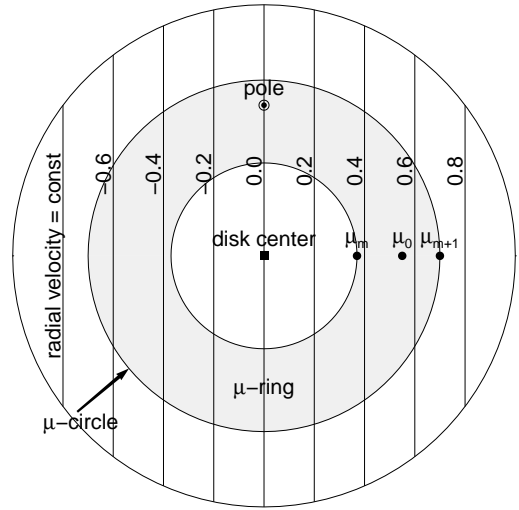}}
\end{center}
\caption[]{Illustration of the apparent radial velocity distribution on a
  stellar disk for solid body rotation: vertical lines of constant radial
  velocity are labeled by their velocity in units of \vsini. They lie
  parallel to rotational pole -- disk center direction.
  For further explanations see text.
\label{f:scheme}}
\end{figure}

We start by describing the procedure of Dravins \&\ Nordlund (1990,
section 2.1) who model rotational broadening as a superposition of line
profiles at various $\mu\equiv\cos(\vartheta)$, broadened by the velocity field
along circles $\mu=\mathrm{const}$ around the stellar disk center. We refer to
one of these circles as ``$\mu$-circle'' (see Fig.~\ref{f:scheme}). The
broadening effect of the projected rotational velocity along a $\mu$-circle
located at $\mu=\mu_\imu$ can be expressed as a convolution according to
\beq
\Irot(v,\mu_\imu) = \frac{1}{\pi}\rotint\frac{\I(v-\xi,\mu_\imu)}{\sqrt{\vsinimu^2-\xi^2}}
\label{e:irotcircle}
\eeq
where we expressed the wavelength-dependence of the intensity
$\I(\lambda,\mu_\imu)$ in the line in terms of the Doppler speed~$v$, $\I(v,\mu_\imu)$.
\Irot\ is the intensity in the line profile broadened under the action of the
radial velocity along the considered $\mu$-circle. \vsinimu\ is the maximum
projected rotational velocity at $\mu_\imu$ given by
\beq
\vsinimu(\mu_\imu) = \vsini \sqrt{1-\mu_\imu^2}. 
\label{e:vsinimu}
\eeq
\vsini\ is the usual projected rotational velocity of the star as a
whole.  An obvious feature of the convolution expressed by
Eq.~\eref{e:irotcircle} is the singularity of the integrand at
$\xi=\pm\vsinimu$, i.e., the presence of sharp spikes. One might already
suspect at this point that their presence will be apparent in the fully
disk-integrated profile. Numerically, one can handle the singularities by
analytically integrating the expression over each resolution element assuming
a certain functional dependence of $\I(v)$ within them, and summing over all
contributions in the interval $\left[-\vsinimu, +\vsinimu\right]$. We assumed a
constant behavior of \I\ in each resolution element so that the integral of
the kernel function over a resolution element stretching over an interval
$\left[\xi_1,\xi_2\right]$ can be conveniently expressed as
\beq
\int^{\xi_2}_{\xi_1}\!\!\frac{d\xi}{\sqrt{\vsinimu^2-\xi^2}}
  =\arcsin\left(\frac{\xi_2}{\vsinimu}\right)-\arcsin\left(\frac{\xi_1}{\vsinimu}\right).
\eeq
The disk-integrated, rotationally-broadened flux profile~$\Frot(v)$ is obtained
by integrating \Irot\ over the disk according to the standard relation
\beq
\Frot(v) = 2\pi\int_0^1\!d\mu\,\mu\Irot(v,\mu) \simeq 2\pi\msum{\imu}\wmu \Irot(v,\mu_\imu).
\label{e:frottotal}
\eeq
The second approximate equality is the discrete approximation to the integral
employing a total number of \nmu\ limb-angles of weight~\wmu\ at positions
$\mu_\imu$. 

The original implementation of Dravins \&\ Nordlund did not use a formulation
as convolution but a discrete integration over $\mu$-circles using polar
coordinates. While mathematically equivalent to Eq.~\eref{e:irotcircle}, it
somewhat obscures the critical role of the most extreme velocities on a
$\mu$-circle for the smoothness of the rotationally broadened spectrum.

\begin{figure}
\resizebox{\hsize}{!}{\includegraphics[draft = \draftflag]%
{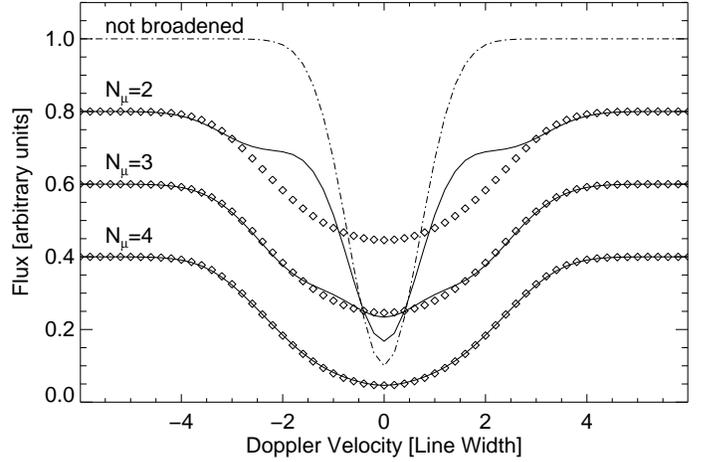}}
\caption[]{Rotationally broadened line profiles (solid lines) for different total numbers of
  asterocentric angles~\nmu\ employing the broadening procedure of Dravins \&\
  Nordlund in comparison to the exact profile (diamonds). For clarity, the
  broadened line profiles have been vertically
  offset. Wavelengths are given as Doppler velocities in units of the width of
  the Gaussian non-broadened profile (dash-dotted line).
\label{f:irot2}}
\end{figure}

Figure~\ref{f:irot2} illustrates the result of a test of Dravins \&\
Nordlund's procedure. We broadened an artificial Gaussian line profile with a
rotational speed~$\vsini$ of three times the line's Doppler width. The
rotational speed was chosen to be particularly critical. Effects of smaller
rotational velocities become less pronounced due to the smoothing by the
Gaussian line profile, at larger rotational speeds deviations are less
conspicuous since less localized. For the test case we assumed that the
relative line shape is independent of $\mu$, and that the intensity in the
continuum follows a linear limb-darkening law with a limb-darkening
coefficient of 0.6. Hence, standard flux convolution can be applied to obtain
the \textit{exact} disk-integrated line profile. Figure~\ref{f:irot2} shows
the convergence of the numerical approximation towards the exact result with
an increasing total number of limb-angles~\nmu. As evident in the figure,
noticeable deviations between the exact and the numerical result are present
up to and including $\nmu=3$.

\section{An improved method}

The major reason for the ``wiggly'' behavior of the broadened line profile
shown in Fig.~\ref{f:irot2} are the pronounced spikes in the integrand in
Eq.~\eref{e:irotcircle}. One can reduce their impact by associating a given
$\I(\mu)$ not only with an infinitely thin $\mu$-circle but with a
\textit{$\mu$-ring} (see Fig.~\ref{f:scheme}) of finite extent. The
contribution~$\Frot(\mu)$ to the rotationally broadened flux profile stemming
from the stellar disk area subtended by $\mu\in\left[\mu_\imu,1\right]$ is
given by the convolution
\beq
\Frot(v,\mu_\imu) = \frac{2}{\vvsini}\rotint\I(v-\xi,\mu_0)\sqrt{\vsinimu^2-\xi^2}.
\label{e:frotring}
\eeq
\vsinimu\ is again given by relation~\eref{e:vsinimu}.  Here, we assume that
the intensity profile is constant within $\left[\mu_\imu,1\right]$ represented
by the intensity at $\mu=\mu_0$ -- presumably but not necessarily lying in the
interval $\left[\mu_\imu,1\right]$ . Note, that relation~\eref{e:frotring}
takes into account the surface area corresponding to the interval
$\left[\mu_\imu,1\right]$. Its result is a flux-like integral, different from
the result of relation~\eref{e:irotcircle} which expresses an average
intensity since the kernel function is normalized to one.  A $\mu$-ring
extending over the interval $\left[\mu_\imu,\mu_\imupone\right]$ can be
obtained by subtracting contributions $\left[\mu_\imupone,1\right]$ from the
contribution of $\left[\mu_\imu,1\right]$ (assuming $\mu_\imupone > \mu_\imu$)
keeping the same intensity at~$\mu_0$.  One can build up the whole visible
stellar disk by a number of $\mu$-rings. Their surface area is reflecting the
integration weight~\wmu\ in Eq.~\eref{e:frottotal}. As stated before, in each
\textit{individual} ring the intensity is assumed to be $\mu$-independent.

\begin{figure}
\resizebox{\hsize}{!}{\includegraphics[draft = \draftflag]%
{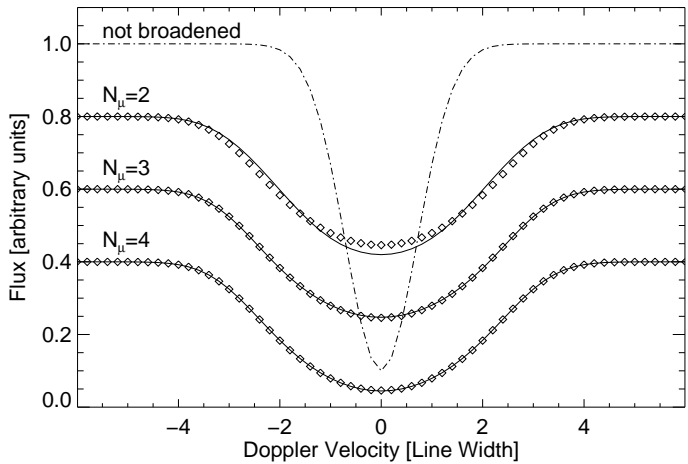}}
\caption[]{As for Fig.~\ref{f:irot2} for our broadening procedure.
  \label{f:irot1}}
\end{figure}

Figure~\ref{f:irot1} illustrates the outcome of this procedure. A comparison with
Fig.~\ref{f:irot2} shows a more rapid convergence towards the exact result.
The improvement is related to the fact that the new method can at least handle
exactly the simple case of a globally $\mu$-independent intensity which is not
the case for the method of Dravins \&\ Nordlund. One could further refine it
by introducing an analytical expression for the $\mu$-dependence of the
intensity in each $\mu$-ring -- perhaps motivated from a fit to the continuum
intensity available at the discrete $\mu_\imu$. However, we did not implement
this since from the test it appeared that the accuracy at an
affordable number of~$\nmu=3$ is already sufficient.  There is a caveat to
this statement: the accuracy of the methods also hinges on the level of the
differential line-shift and -broadening as a function of $\mu$ which we did
not test here. This should be checked on a case by case basis.

\section{Remarks on differential rotation}

Up to this point we considered solid body rotation only. However, the previous
discussion made clear that integration over $\mu$-rings should also perform
better in the case of differential rotation. The referee made the point that
putting efforts into 3D model and spectral synthesis calculations warrants the
inclusion of effects of differential rotation to maintain highest level of accuracy.
In the following we give a brief demonstration that differential rotation on
the level observed in the Sun can indeed be relevant for resulting line
profiles. We implemented our method for a case of differential rotation.

We chose the possibly simplest parameterization of the solar
differential rotation pattern of the form
\beq
\Omega = \Omega_\mathrm{eq} \left(1-\alpha\cos^2\psi\right) .
\label{e:difrotpattern}
\eeq 
$\Omega$ is the angular velocity at co-latitude~$\psi$, $\Omega_\mathrm{eq}$
the equatorial angular velocity, and $\alpha$ a dimensionless parameter
measuring the degree of differential rotation.  $\alpha\approx0.2$ for the Sun
(see, e.g., Reiners \&\ Schmitt 2002). Knowing the stellar radius, the angular
velocity can be transformed into a relation for the radial velocity as a
function of position on the stellar disk.  However, its complex functional
dependence does not allow the expression of the rotation kernel by elementary
functions analog to Eq.~\eref{e:frotring} (Huang 1961). We evaluated the
kernels for individual $\mu$-rings numerically by discretizing the stellar
disk employing polar coordinates.  While seemingly straight-forward it proved
difficult to obtain sufficient numerical accuracy, in particular at low radial
velocities, and higher-than-expected resolution was necessary. We ensured
numerically that the symmetry of the kernel functions with respect to the
origin (i.e., zero radial velocity) was maintained so that no artificial line
asymmetries were introduced in the resulting line profiles.

\begin{figure}
\resizebox{\hsize}{!}{\includegraphics[draft = \draftflag]%
{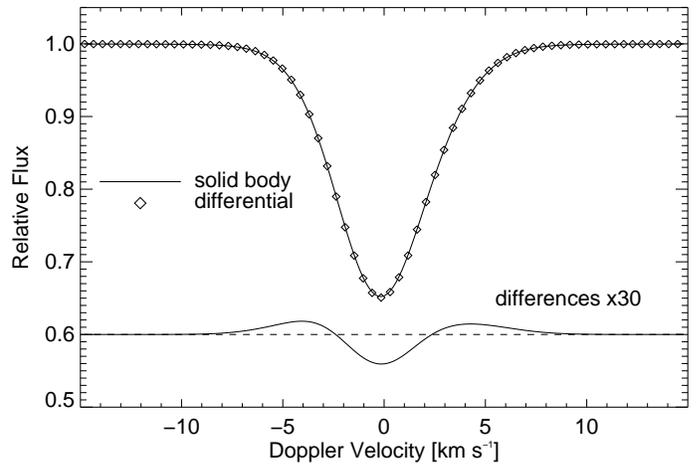}}
\caption[]{Comparison of rotational broadening assuming a rigidly (solid line) 
  and differentially (diamonds) rotating Sun. Thirty times enlarged differences of
  the resulting profiles (in the sense differential--rigid) are plotted in
  the lower part of the panel. For further explanations see text. 
  \label{f:drot}}
\end{figure}

Figure~\ref{f:drot} depicts an example comparing rotationally broadened
profiles assuming rigid as well as differential rotation. We arbitrarily
selected an \ion{Fe}{I} line (at 6082\,\AA) of a 3D spectral synthesis
calculation for the Sun, and solar-like rotational parameters,
$V=1.8$\pun{km\,s$^{-1}$}, $\sin(i)=1.0$, and for the case of differential
rotation $\alpha=0.2$.  The plot shows that -- as expected -- our procedure of
integrating over $\mu$-circles (here $N_\mu=4$) leaves no obvious imprint of
spikes in the resulting profiles. Moreover, noticeable differences are present
for {\em fixed $V$ and $\sin(i)$}. One might interpret the smaller degree of
broadening in the differentially rotating case as simply a result of the
the smaller disk-averaged rotation rate for $\alpha > 0$. However,
from Eq.~\eref{e:difrotpattern} we obtain for the root-mean-square radial
velocity~$v_\mathrm{RMS}$ due to the rotation over the stellar disk
\beq
v_\mathrm{RMS}=\sqrt{\frac{2}{3}} V\sin(i)
\sqrt{1-\alpha\left(\frac{3}{4}\cos(i) + \frac{1}{5}\alpha\right)}.
\label{e:vradrms}
\eeq 
For our test case Eq.~\eref{e:vradrms} gives a 0.4\% smaller
$v_\mathrm{RMS}$ when $\alpha=0.2$ instead of zero. Changing the equatorial
velocity~$V$ by this amount assuming rigid rotation, leads to much smaller
changes in the line profile than are visible in Fig.~\ref{f:drot}. However, for
the particular Fe line it is possible to closely emulate its profile for
differential rotation assuming rigid rotation where $V$ is 3\% smaller than
its nominal value.

\section{Conclusions}

The approach suggested in this paper provides an accurate treatment of
rotational line broadening of a spherical star employing intensities at
a modest number of limb-angles only. This is of practical importance in the
context of 3D spectral synthesis calculations where angular resolution is
computationally expensive. This also holds for differential rotation but
  the rotational kernel functions must be evaluated numerically.  Our
approach is not restricted to the 3D case but could be equally well applied in
the standard 1D case. The issue is of course less pertinent in 1D since high
angular resolution is affordable making the broadening method uncritical.

\acknowledgement We thank Ansgar Reiners for helpful discussions about
differential rotation. The work was funded by EU grant MEXT-CT-2004-014265
(CIFIST).

\end{document}